\documentclass[final,5p,times,twocolumn]{elsarticle}

\usepackage{amssymb}
\usepackage{amsmath}
\usepackage{lipsum}
\usepackage{graphicx}
\usepackage{dcolumn}
\usepackage{bm}
\usepackage{braket}
\usepackage{epstopdf}
\usepackage{hyperref}

\usepackage{color}
\usepackage{cancel}
\usepackage{subfigure}

\biboptions{sort&compress}

\journal{Physics Letters B}

\begin{document}

\begin{frontmatter}



\title{Nucleon-charmonium interactions from lattice QCD}
\author[ithems]{Yan Lyu}
\ead{yan.lyu@riken.jp}
\author[ithems]{Takumi Doi}
\ead{doi@ribf.riken.jp}
\author[ithems]{Tetsuo Hatsuda}
\ead{thatsuda@riken.jp}
\author[Rissho,ithems]{Takuya Sugiura}
\ead{sugiura@rcnp.osaka-u.ac.jp}
\affiliation[ithems]{organization={Interdisciplinary Theoretical and Mathematical Sciences Program (iTHEMS), RIKEN}, city={Wako}, postcode={351-0198}, country={Japan} }
\affiliation[Rissho]{organization={Faculty of Date Science, Rissho University}, city={Kumagaya}, postcode={360-0194}, country={Japan}}

\begin{abstract}
We present a realistic lattice QCD study on low-energy $N$-$J/\psi$ and $N$-$\eta_c$ interactions
based on (2+1) flavor configurations with nearly physical pion mass $m_\pi=146$ MeV.
The interactions, extracted from the spacetime correlations of nucleon and charmonium system by using the HAL QCD method,
are found to be attractive in all distances and possess a characteristic long-range tail consistent with the two-pion exchange potential. 
The resulting $S$-wave scattering lengths are $0.30(2)\left(^{+0}_{-2}\right)$ fm, $0.38(4)\left(^{+0}_{-3}\right)$ fm, and $0.21(2)\left(^{+0}_{-1}\right)$ fm
for spin-$3/2$ $N$-$J/\psi$, spin-$1/2$ $N$-$J/\psi$, and spin-$1/2$ $N$-$\eta_c$, respectively. Our results are orders of magnitude larger than those from the photoproduction experiments assuming the vector meson dominance. 
Our findings may provide deeper understanding of the nonperturbative QCD phenomena 
 ranging from the origin of nucleon mass to the in-medium 
  $J/\psi$ mass modification as well as the properties of hidden-charm pentaquark states.
\end{abstract}


\end{frontmatter}

\section{Introduction}\label{introduction}
Quantum Chromodynamics (QCD) governs not only the interaction among quarks and gluons at short distances but also the interaction between color-neutral hadrons at long distances, and thus is responsible for perturbative and non-perturbative hadronic phenomena of the strong interaction.
In particular, questions such as how hadrons gain (or lose) mass in the vacuum (medium), and what kind of exotic hadrons other than the usual mesons and baryons can exist, have attracted much attention in recent years.

Among others, the low-energy interaction between a nucleon ($N$) and a charmonium 
($J/\psi$ and $\eta_c$) has intimate connections with the above two questions. 
On the one side, 
since $N$ and $c\bar{c}$ do not have common valence quarks,
the compact $c\bar{c}$ is expected to interact with $N$ mainly through
multiple-gluon exchange at low energies. Consequently, 
the $N$-$c\bar{c}$ forward scattering amplitude is related to the nucleon matrix element of gluon field~\cite{Kharzeev:1995ij}, and which in turn is connected to the trace anomaly contribution to the nucleon mass~\cite{Shifman:1978zn,Ji:1994av}.
Determining the gluon structure of the nucleon is also one of the major goals of the GlueX experiment~\cite{GlueX:2019mkq} at JLab and future electron-ion colliders~\cite{Accardi:2012qut,Anderle:2021wcy}.
In addition, the $N$-$J/\psi$ scattering length gives an access to the $J/\psi$ mass modification in nuclear medium~\cite{Hayashigaki:1998ey}, whose intimate relation to the in-medium 
QCD condensations has been extensively discussed~\cite{Hayano-Hatsuda2010,Gubler:2018ctz}.

On the other side, exploring multiquark hadrons has attracted enormous experimental and theoretical interests for decades~\cite{Chen2016,Guo2017,Olsen2017,Brambilla2019}. The first pentaquark candidate $P_c$ was observed in $p$-$J/\psi$ spectrum by the LHCb Collaboration~\cite{LHCb2015_Pc,LHCb:2019kea}.
A complete understanding on its nature and properties needs a reliable $N$-$c\bar{c}$ interaction, which serves as an indispensable ingredient in a full coupled-channel analysis of $P_c$.
Moreover, the possible existence of charmonium-nucleus bound states ($c\bar{c}$-$A$) has been widely discussed (see Refs.~\cite{Brodsky1990,Wu-Lee-2011, Yokota:2013sfa,Krein:2017usp} and references therein).
Despite the general agreement on the existence of such bound states among theories, predicted binding energies are quite disparate due to the large uncertainty of $N$-$c\bar{c}$ interaction.
Therefore, a reliable prediction on $c\bar{c}$-$A$ bound states also calls for an accurate $N$-$c\bar{c}$ interaction.
Finally, $N$-$c\bar{c}$ interaction contributes to the $N$-$J/\psi$ total cross section, relevant to the studies of the $J/\psi$ suppression~\cite{Matsui:1986dk} and the intrinsic charm of the nucleon~\cite{Brodsky:1980pb}.

The low-energy interaction between a nucleon and a charmonium ($N$-$c\bar{c}$) is also theoretically interesting in its own right. The interaction, which arises from multiple-gluon exchange, closely resembles the van der Waals force \cite{Bhanot:1979vb}. However, unlike the case of the van der Waals interaction in QED, the multi-gluon exchange in QCD 
is necessarily non-perturbative and is likely dominated by the OZI-violating two-pion exchange at long distances.\cite{Fujii:1999xn,Brambilla:2015rqa,Castella2018}. Therefore, it is important to investigate this hypothesis through first-principles QCD calculations. 

Experimentally, the low-energy $N$-$c\bar{c}$ interaction, however, is
poorly constrained due to the lack of a direct measurement of $N$-$c
\bar{c}$ two-body system. For instance, the $N$-$J/\psi $ scattering lengths
inferred from analyzing data of the $J/\psi $ photoproduction off the proton
with the vector meson dominance and with the low-energy unitarity are
$O(1\sim 10)\times 10^{-3}$~fm~\cite{Pentchev:2020kao} and $O(1)$ fm~\cite{JPAC:2023qgg},
respectively. Pioneering lattice studies are limited by quenched approximation~\cite{Yokokawa:2006td,Kawanai:2010ru,Kawanai:2010ev},
the use of heavy pion mass~\cite{Sugiura:2019pye}, and/or large uncertainties
leading to scattering length effectively zero~\cite{Liu:2008rza,Skerbis:2018lew}.

Under these circumstances, the purpose of this Letter is to present a realistic study on the low-energy $N$-$c\bar{c}$ interaction based on first-principles lattice QCD calculations with nearly physical quarks masses.
As we shall see below, the HAL QCD method, which converts the hadronic spacetime correlation function to physical observables via a potential, provides us with a suitable theoretical tool for identifying the characteristic long-range $N$-$c\bar{c}$ potential and accurately determining various scattering properties.

\section{HAL QCD Method} \label{HAL}
Let us start with the following equal-time Nambu-Bethe-Salpeter (NBS) amplitude,
\begin{align}\label{Eq_NBS}
    \psi_E(\bm r)e^{-Et} = \sum_{\bm x} \braket{0|N(\bm x+\bm r, t)O_{c\bar c}(\bm x, t)|N,c\bar c;E},
\end{align}
where $N$ and $O_{c\bar c}$ are the nucleon and charmonium operators, respectively. $\ket{N,c\bar c;E}$ is an eigenstate of $N$-$c\bar c$ system with total energy $E=\sqrt{m^2_N+\bm k^2}+\sqrt{m^2_{c\bar c}+\bm k^2}$, $\bm k$ being the momentum in the center of mass frame.
It has been shown that the scattering phase shift is encoded into the above NBS amplitude~\cite{Ishii2007,Aoki2020}.
Thanks to the Haag-Nishijima-Zimmermann reduction formula for composite particles~\cite{Zimmermann1987}, 
the interaction between hadrons can be defined as an energy-independent nonlocal potential $U(\bm r, \bm r')$ as follows~\cite{Ishii2007,Aoki2010}, 
\begin{align}\label{Eq_HAL}
    (E_k-H_0)\psi_E(\bm r)=\int d\bm r' U(\bm r, \bm r')\psi_E(\bm r'),
\end{align}
where $E_k=\frac{\bm k^2}{2\mu}$ and $H_0=-\frac{\nabla^2}{2\mu}$ with $\mu$ being the reduced mass.

In many cases, instead of a NBS amplitude, 
the following normalized correlation function $R(\bm r,t)$ consisting of several NBS amplitudes with difference energies 
is much more easier to be obtained in practical lattice simulations, 
\begin{align}\label{Eq_R}
    R(\bm r,t)&=\sum_{\bm x}\braket{0|N(\bm x+\bm r, t)O_{c\bar c}(\bm x, t)\overline{\mathcal{J}}(0)|0}/e^{-(m_N+m_{c\bar c})t}\\ \nonumber
    &=\sum_{n}a_n\psi_{E_n}(\bm r)e^{-(\Delta E_n)t} + O(e^{-(\Delta E^*)t}).
\end{align}
Here $\mathcal{J}(0)$ is a source operator, the overlapping factor $a_n$ is defined as $a_n=\braket{N,c\bar c;E_n|\overline{\mathcal{J}}(0)|0}$. $\Delta E_n=E_n-(m_N+m_{c\bar c})$ is the energy of the $n$th eigenstate with respect to the $N$-$c\bar c$ threshold, $\Delta E^*\sim m_\pi$ is the inelastic threshold.
As shown in Ref.~\cite{Ishii2012}, $R(\bm r, t)$ at large $t$ satisfies the following integrodifferential equation,
\begin{align}
    &\left[\frac{1+3\delta^2}{8\mu}\frac{\partial^2}{\partial t^2}-\frac{\partial}{\partial t} - H_0 +O(\delta^2\partial^3_t)\right] R(\bm r,t)\\ \nonumber
    &=\int d\bm r' U(\bm r, \bm r')R(\bm r',t),
\end{align}
with $\delta=(m_N-m_{c\bar c})/(m_N+m_{c\bar c})$ being the mass asymmetry. $O(\delta^2\partial^3_t)$ term is found to be consistent with zero within statistical uncertainties, and is neglected in our study.
In practice, it is useful to expand the nonlocal potential in terms of nonlocality~\cite{OKubo-Marshak1958,Iritani2019PRD},
$U(\bm r,\bm r')=V(r)\delta(\bm r-\bm r')+\sum_{n=1}V_n(\bm r)\nabla^n\delta(\bm r-\bm r')$. A leading-order (LO) potential is accurate enough for the near-threshold scattering, and reads,
\begin{align}\label{Eq_LO_V}
    V(r)=R^{-1}(\bm r,t)\left[\frac{1+3\delta^2}{8\mu}\frac{\partial^2}{\partial t^2}-\frac{\partial}{\partial t} - H_0 \right] R(\bm r,t).
\end{align}
The truncation error from higher-order terms of the derivative expansion 
can be estimated through the $t$ dependence of $V(r)$~\cite{Iritani2019Jhep,Lyu2022}.

\section{Lattice setup}\label{LQCD-setup}
\subsection{Interpolating Operators}
We use local sink operators, defined as,
\begin{align}
    N_\alpha(x)  &= \epsilon_{ijk}[u^i(x)C\gamma_5 d^j(x)]u^k_\alpha(x), \label{Eq_N_op} \\ 
    J/\psi_\mu(x) &= \delta_{ij} \bar c^i(x)\gamma_\mu c^j(x), \label{Eq_Jpsi_op} \\ 
    \eta_c(x) &= \delta_{ij} \bar c^i(x)\gamma_5 c^j(x), \label{Eq_etac_op}
\end{align}
where $i$, $j$, and $k$ are color indices, $\alpha$ ($\mu$) is spinor (vector) index.
$C=\gamma_4\gamma_2$ is the charge conjugation.
For the source operator, the wall-type smearing with Coulomb gauge fixing is adopted, where quark field is projected to the zero-momentum mode to enlarge the overlap with low-energy $N$-$c\bar c$ states. Explicit forms are given by making the replacement $q(\bm x, t)\rightarrow \mathcal{Q}(t)=\sum_{\bm x}q(\bm x, t)$ for all quark operators in Eqs.~(\ref{Eq_N_op})-(\ref{Eq_etac_op}).

The spin projection is performed for both sink and source operators. For instances, $N$-$J/\psi$ sink operators with $(s,s_z)=(1/2, + 1/2)$ and $(s,s_z)=(3/2, +1/2)$ can be constructed as,
\begin{align}
    &[NJ/\psi]_{\frac{1}{2},+\frac{1}{2}}=\frac{-1}{\sqrt{3}}(J/\psi_1- iJ/\psi_2)N_2-\frac{1}{\sqrt{3}}J/\psi_3N_1, \\
    &[NJ/\psi]_{\frac{3}{2},+\frac{1}{2}}=\frac{-1}{\sqrt{6}}(J/\psi_1- iJ/\psi_2)N_2+\sqrt{\frac{2}{3}}J/\psi_3N_1,    
\end{align}
with $N_{1,2}$ and $J/\psi_{1,2,3}$ defined in Eqs. (\ref{Eq_N_op})-(\ref{Eq_Jpsi_op}).
Finally, the Misner’s method for approximate partial wave decomposition on a cubic grid, originally proposed in numerical general relativity~\cite{Misner:1999ab}, is used to perform $S$-wave projection on the lattice~\cite{Miyamoto2020}.

\subsection{Lattice QCD Configurations}
The ($2+1$)-flavor gauge configurations are generated on a $L^4=96^4$ lattice volume with lattice spacing $a\simeq 0.0846$ fm ($a^{-1}\simeq2333$ MeV), which corresponds to $La\simeq8.1$ fm.
The Iwasaki gauge action at $\beta=1.82$ and the nonperturbatively $O(a)$-improved Wilson quark action with stout smearing at nearly physical quark masses are adopted~\cite{Ishikawa2016}.
For the charm quark, the relativistic heavy quark (RHQ) action is used to remove the cutoff errors associated with the charm quark mass up to next-to-next-to-leading order~\cite{Aoki2003}.
With two sets (set 1 and set 2) of RHQ parameters~\cite{Namekawa2017} tuned to be very close to the physical charm quark mass, we are able to interpolate ($0.385~\times$ set 1 $+$ $0.615~\times$ set 2) the physical charm quark mass and reproduce the dispersion relation for the spin-averaged $1S$ charmonium.

\subsection{Measurement and statistics}
We perform 80 measurements by shifting the source position in a temporal direction for each
of 200 gauge configurations separated by 10 trajectories.
To further reduce the statistical fluctuation, the forward and backward propagations are averaged in each configuration, the hypercubic symmetry on the lattice (four
rotations) is utilized, which lead to 128,000 measurements in total.
The domain-decomposed solver~\cite{Ishikawa2023} and the Bridge++ code~\cite{Bridge} with the periodic boundary condition for all directions are used to calculate the light and charm quark propagators, respectively.
The hadronic correlation functions are obtained by the unified contraction algorithm~\cite{Doi2013}.
Throughout this paper, we adopt the jackknife method with a bin size of 20 configurations to calculate the statistical errors. A comparison with a bin size of 40 configurations shows that the bin size dependence is small.

Shown in Table~\ref{tab-mass} are isospin-averaged masses of relevant hadrons calculated from our lattice simulations together with the corresponding experimental values.
The lattice results, obtained by a single-state fit in the interval $t/a=15-25$ for $\pi$ and $K$, $11-18$ for $N$, and $30-40$ for $J/\psi$ and $\eta_c$, are just few percent deviation from the experimental results.

\begin{table}[htbp]
\begin{center}
\caption{ Isospin-averaged hadron masses with statistical errors obtained from (2+1)-flavor lattice QCD simulations together with the experimental values.
Two values for $J/\psi$ and $\eta_c$ are from set 1 and set 2 parameters of RHQ action, respectively.
}
\begin{tabular}{ccc}
  \hline\hline
    Hadron &~~Lattice [MeV]  &~~Expt. [MeV] \\
  \hline
    $\pi$ &\ \ 146.4(4)  & \ \ 138.0\\
    $K$  & \ \ 524.7(2) & \ \ 495.6\\
    $N$  &\ \ 954.0(2.9)  & \ \ 938.9 \\
    $J/\psi$ &\ \ 3121.1(1)~~3076.3(1) &\ \ 3096.9\\
    $\eta_c$ &\ \ 3022.8(1)~~2976.3(1)&\ \ 2984.1\\
  \hline\hline
\end{tabular} \label{tab-mass}
\end{center}
\end{table}
\section{Nucleon-charmonium interactions}

\subsection{Potentials in different channels}
\begin{figure*}[htbp]
    \centering
    \subfigure[]{\includegraphics[width=8.0cm]{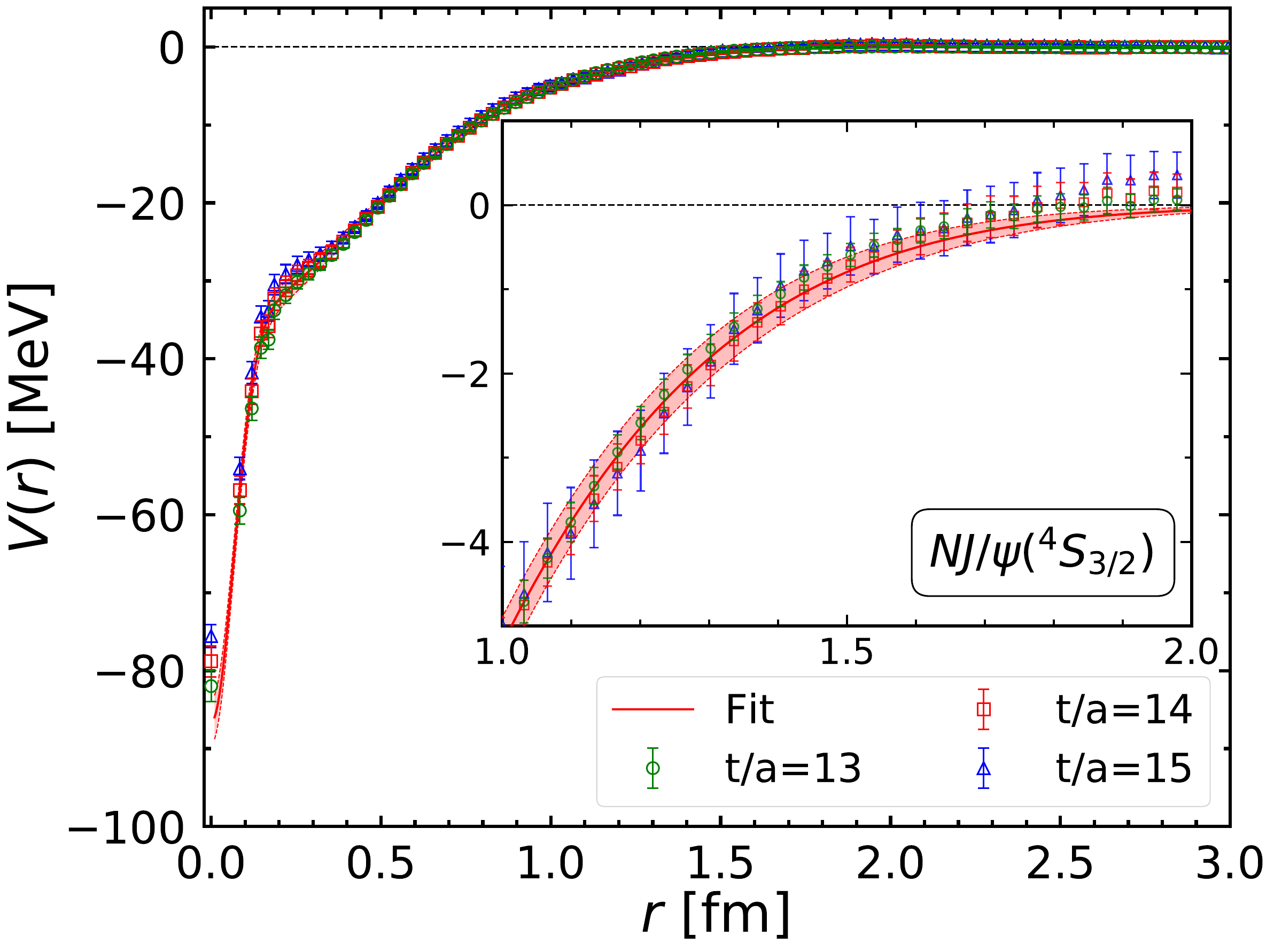}}
    \subfigure[]{\includegraphics[width=8.0cm]{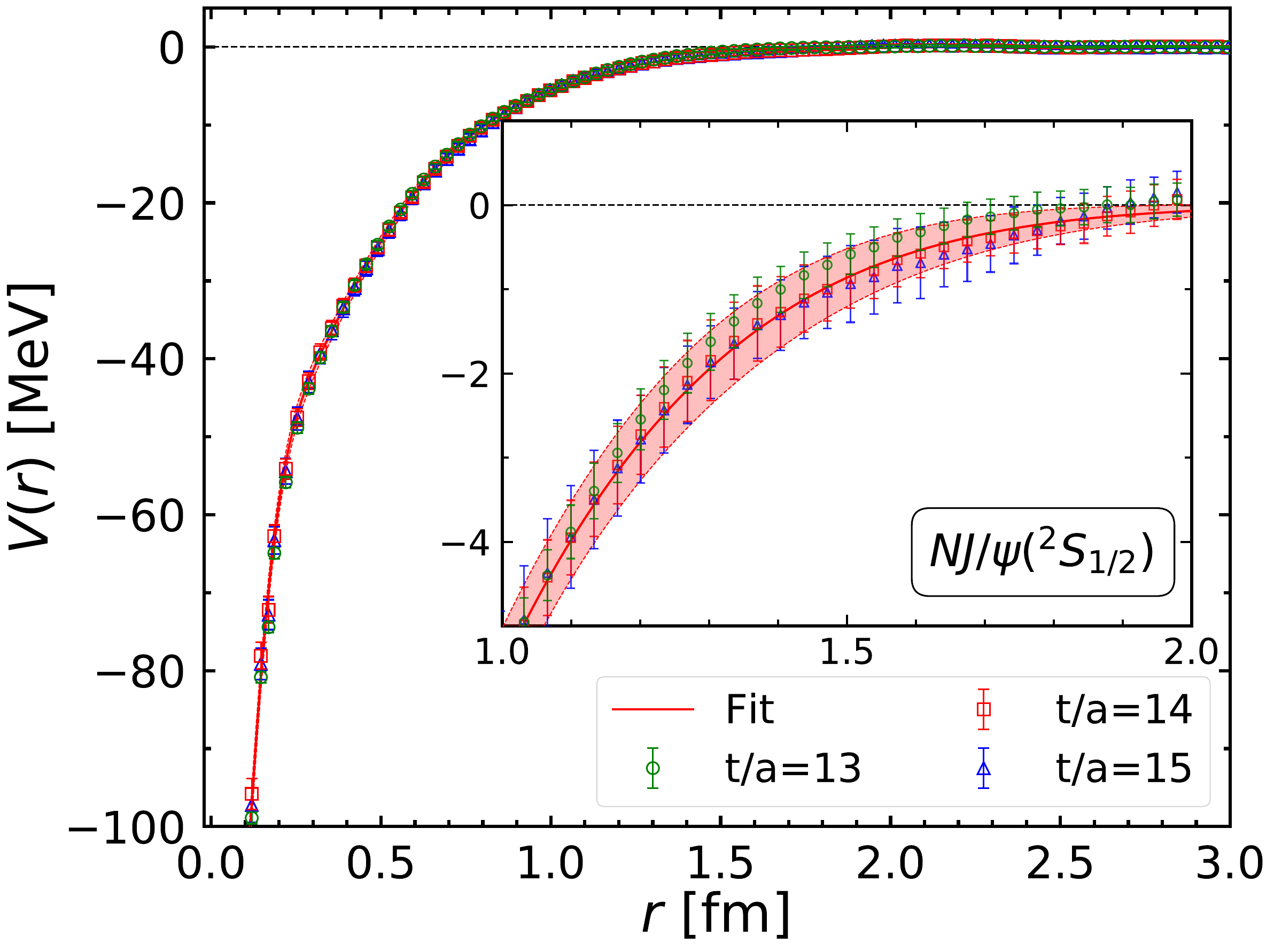}}
    \subfigure[]{\includegraphics[width=8.0cm]{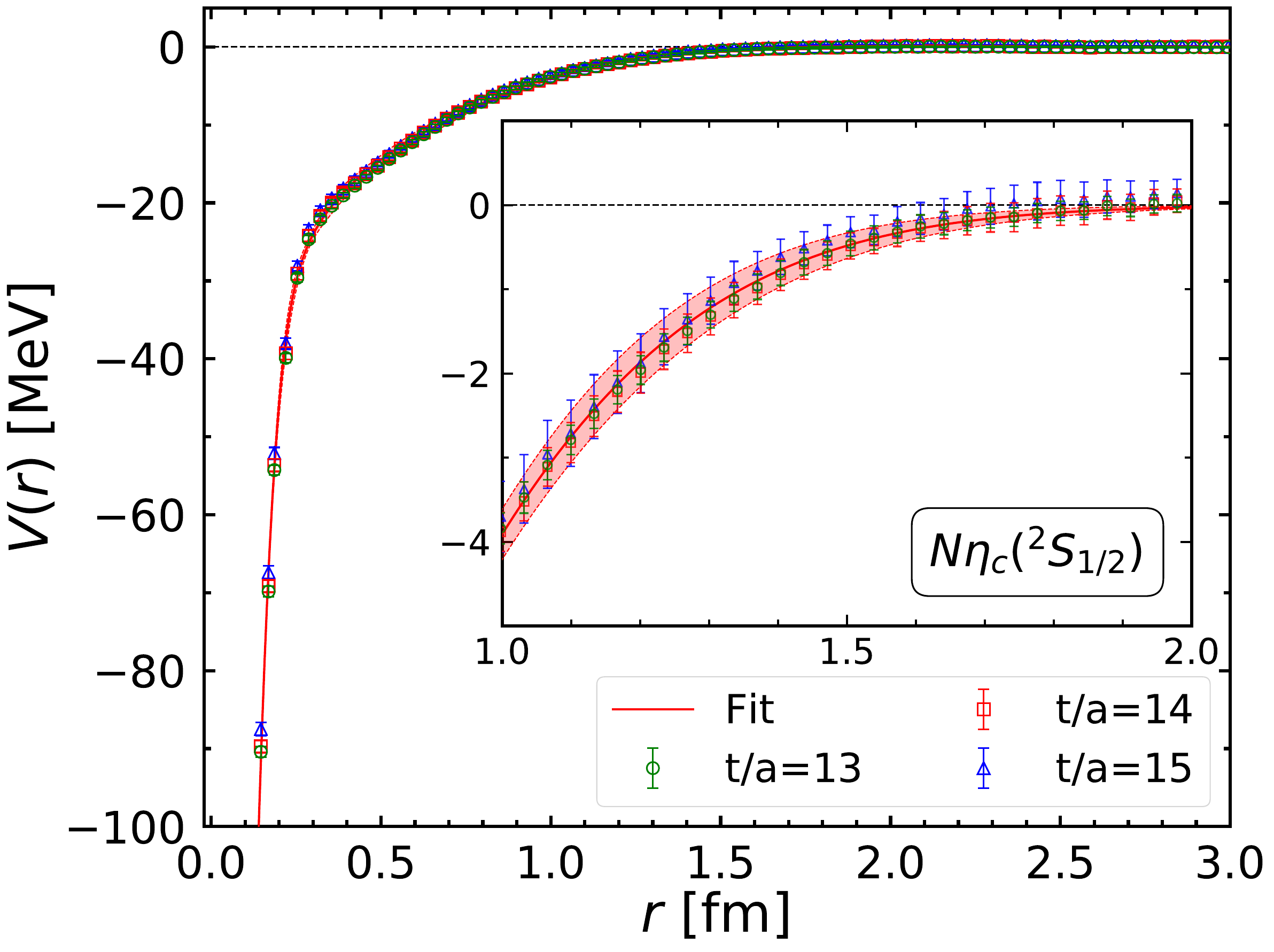}}
    \subfigure[]{\includegraphics[width=8.0cm]{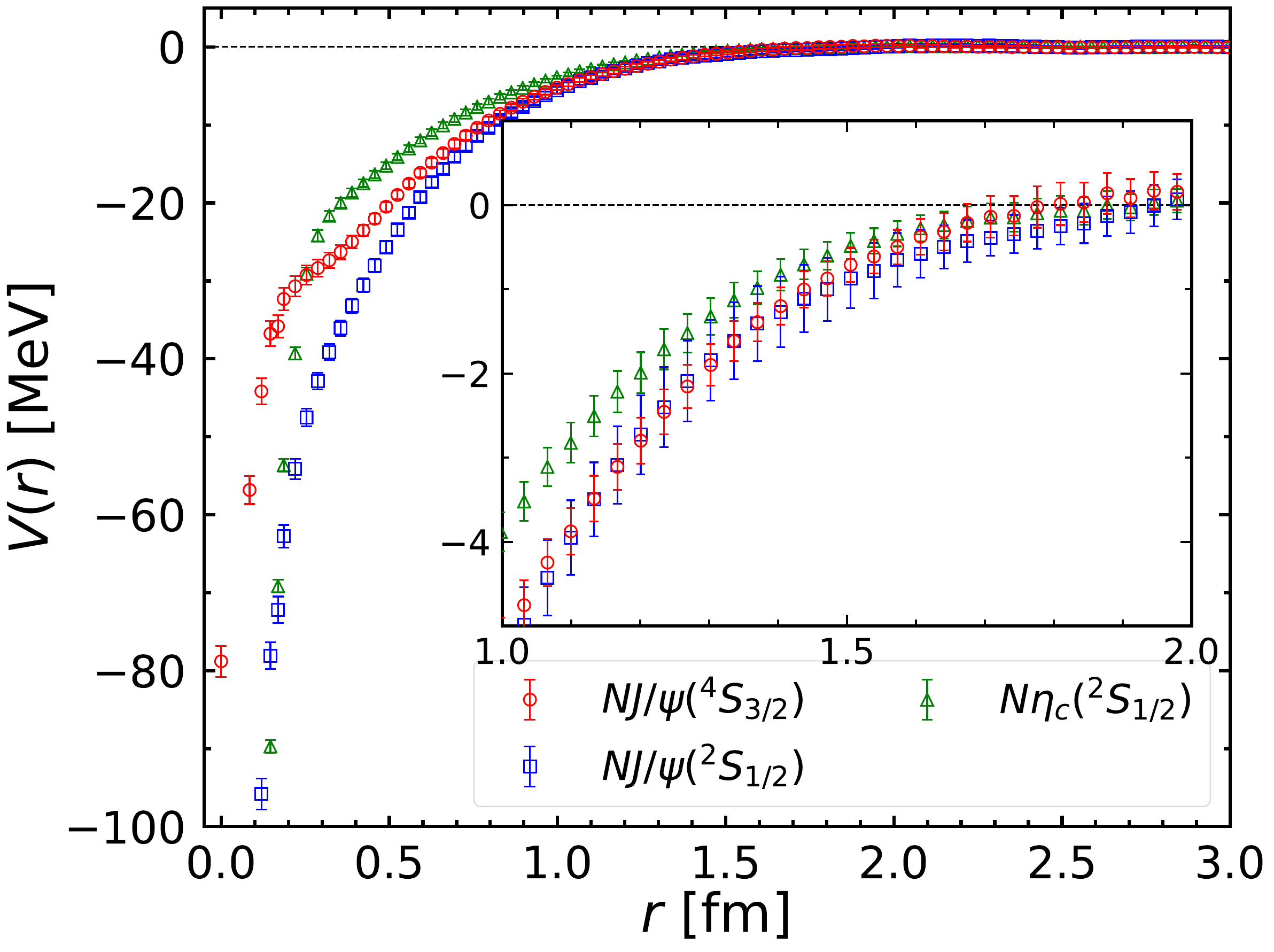}}
    \caption{The $S$-wave $N$-$c\bar c$ potential extracted at $t/a=13$, $14$, and $15$ for $N$-$J/\psi$ with $^4S_{3/2}$ (a), with $^2S_{1/2}$ (b), and $N$-$\eta_c$ with $^2S_{1/2}$ (c). The red bands show the fit results with phenomenological three-range Gaussians at $t/a=14$.
    The three potentials at $t/a=14$ are also shown in (d) for a direct comparison.
    A magnification is shown in the inset for each panel.
    }
    \label{Fig-V}
\end{figure*}

With the lattice measurements of the normalized correlation function $R(\bm r, t)$ in Eq.~(\ref{Eq_R}), we are ready to calculate the LO potential in Eq.~(\ref{Eq_LO_V}).
Shown in Fig.~\ref{Fig-V} are the $S$-wave $N$-$c\bar c$ potentials from the interpolation between set 1 and set 2
for $N$-$J/\psi$ with $^{4}S_{3/2}$ (a), with $^{2}S_{1/2}$ (b), and for $N$-$\eta_c$ with $^{2}S_{1/2}$ (c).
(The $^{2s+1}L_J$ notation is used with total spin $s$, orbital and total angular momenta $L$ and $J$.)
The systematic uncertainty associated with the interpolation is negligible compared to the current statistical error, as the potentials from set 1 and set 2 are found to be nearly identical.
The three potentials are extracted at $t/a=13$, $14$, and $15$, corresponding to $t\simeq1.2$ fm,
which is large enough to suppress inelastic excited-state contaminations at smaller $t$, and at the same time is small enough to avoid  the exponentially increasing 
statistical fluctuations at larger $t$ associated with the nucleon propagation~\cite{Parisi:1983ae,Lepage1989}.
Potentials in Fig.~\ref{Fig-V} are found to be weakly dependent on $t$, suggesting that the systematic error originated from the inelastic states such as $\Lambda_c \bar{D}^{(*)}$ and the truncation of the derivative expansion is small.
Such small variations over $t$ are included as systematic error in our final results.
We recall that the transition from $N$-$J/\psi$ to $N$-$\eta_c$
is highly suppressed by the heavy-quark spin symmetry.
As we focus on the near-threshold $N$-$c\bar c$ scattering, single-channel calculations in our present study are well justified.

All three potentials in Fig.~\ref{Fig-V} are attractive for all distances,
and possess a characteristic two-component structure: the attractive core at short distances
and the attractive tail at long distances. 
Unlike the nucleon interaction where a repulsive core appears,
here the absence of a repulsive core in the $N$-$c\bar c$ interaction is partially attributed to the fact that the nucleon $N$ and the charmonium $c\bar c$ do not have common valence quarks so that the Pauli exclusion principle between common quarks does not operate.
Indeed, similar short-range attractions have been also observed for other hadron pairs in which the two hadrons do not have common valence quarks, such as $N$-$\phi$~\cite{Lyu_Nphi_PRD2022}, and $N$-$\Omega$~\cite{Iritani2019PLB}.
Fig.~\ref{Fig-V} (d) shows a comparison among $N$-$J/\psi(^4S_{3/2})$, $N$-$J/\psi(^2S_{1/2})$, and $N$-$\eta_c(^2S_{1/2})$ potentials extracted at $t/a=14$.
$N$-$J/\psi$ potential is found to be slightly stronger than $N$-$\eta_c$ potential at short range, consistent with previous observations found by lattice calculations under the quenched approximation~\cite{Kawanai:2010ev} and at heavy pion mass~\cite{Sugiura:2019pye}.

On the other side, we find that $N$-$J/\psi(^4S_{3/2})$, $N$-$J/\psi(^2S_{1/2})$, and $N$-$\eta_c(^2S_{1/2})$ potentials have a similar long-range behavior, and they become nearly degenerate at large distances, indicating a common underlying mechanism dictates these three interactions at large distances.

\subsection{Long-range two-pion exchange}
As mentioned in the Introduction, the low-energy interaction between a nucleon and a heavy quarkonium ($Q\bar{Q}$) is expected to be dominated by multiple-gluon exchange.
However, gluons cannot propagate over large distances because of color confinement.
As a result, the dominant degrees of freedom at long distances involve the color-neutral states coupling to gluons. In particular, the lightest two-pion state would dominate over other states~\cite{Bhanot:1979vb,Fujii:1999xn,Brambilla:2015rqa}.
As has been shown in Refs.~\cite{Castella2018}, the two-pion-exchange (TPE) interaction for $N$-$Q\bar{Q}$ takes the following form in the coordinate space,
\begin{align}
    V(r\gg 1/(2m_\pi))=-\alpha \frac{e^{-2m_\pi r}}{r^2}, \label{Eq_TPE}
\end{align}
with a potential strength $\alpha$.
This is radically different from the QED van der Waals interaction, where two photons can propagate over large distances, leading to a power-law fall-off potential $V(r)\sim-1/r^n$ with $n=6$
 (the van der Waals force without the retardation effect) or $n=7$
 (the Casimir-Polder force with the retardation effect).

\begin{figure}[htbp]
    \centering
    \includegraphics[width=8.0cm]{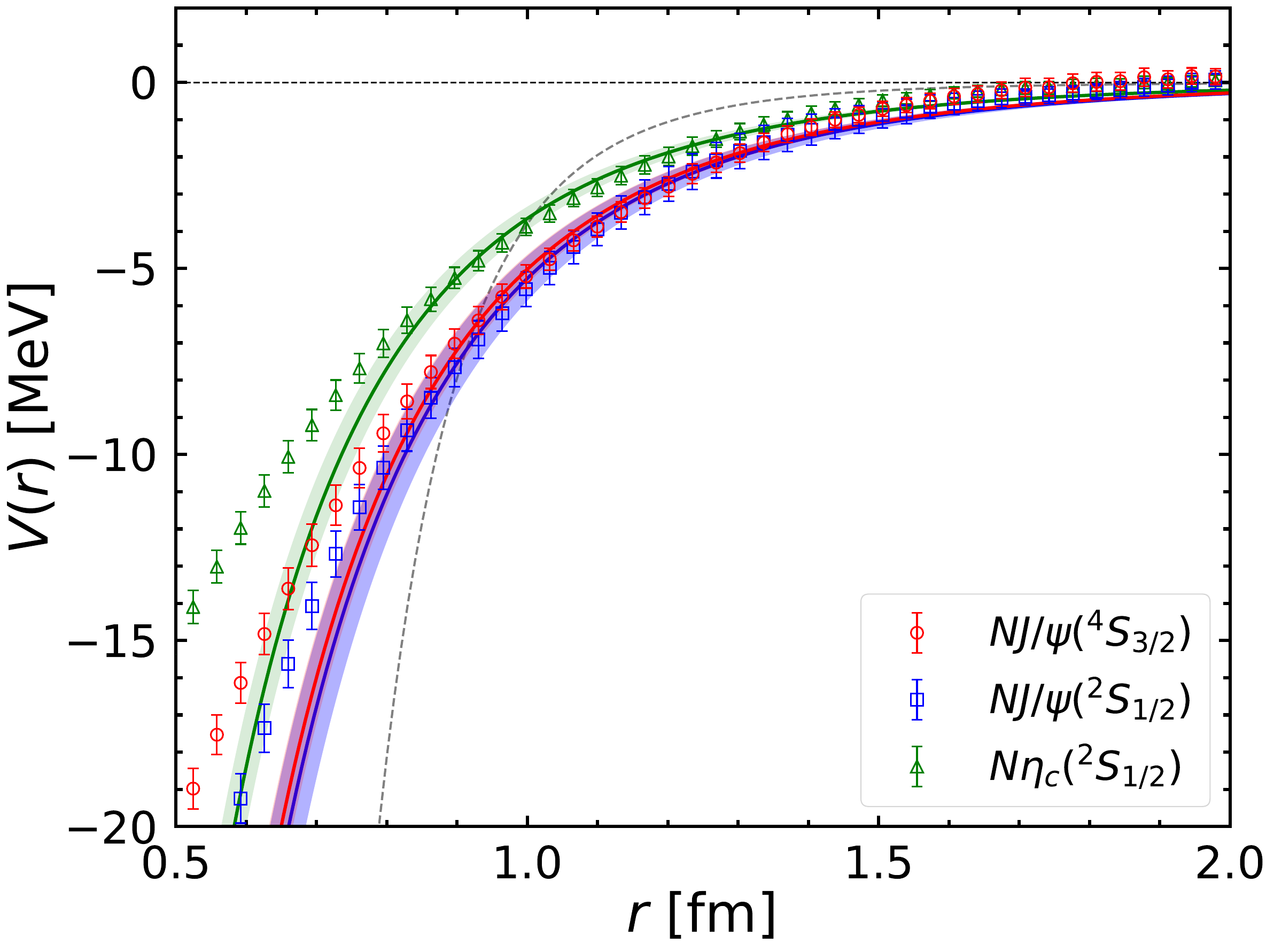}
    \caption{The bands show the fit with the TPE function $V(r)=-\alpha e^{-2m_\pi r}/r^2$ to the long-range $N$-$c\bar{c}$ potential. The gray dashed line is the best fit with $V(r)=-\alpha/r^7$ for comparison.
    }
    \label{Fig-V-TPE}
\end{figure}

In order to check such a long-range behavior, we fit lattice data of the long-range potential with the TPE function in Eq.(\ref{Eq_TPE}).
The fit range is $r_\text{min}< r< r_\text{max}$ with $r_\text{max}=1.8$ fm beyond which the potential vanishes within statistical uncertainties.
By increasing $r_\text{min}$ from $1/2m_\pi\simeq0.7$ fm, we find 
the lattice data in the range of $0.9< r< 1.8~\text{fm}$
can be decently described by Eq.(\ref{Eq_TPE}) with a fit parameter $\alpha=22(2)$, $23(3)$, and $16(2)$ $\text{MeV}\text{fm}^2$ for $N$-$J/\psi$ with spin $3/2$, with spin $1/2$, and $N$-$\eta_c$, respectively.
Shown in Fig.~\ref{Fig-V-TPE} by bands are corresponding fit results.
The best fit with $V(r)=-\alpha/r^7$ in the same fit range is also shown in Fig.~\ref{Fig-V-TPE} by the gray dashed line.
These results imply that the long-range $N$-$c\bar{c}$ interaction is consistent with the TPE potential.

\section{Scattering properties}
\subsection{Phase shifts and scattering parameters}
In order to convert the $N$-$c\bar c$ potential in Fig.~\ref{Fig-V} to its scattering properties, we perform an uncorrelated fit with phenomenological three-range Gaussians 
$V_\text{fit}(r)=-\sum_{i=1}^3 a_ie^{(-r/b_i)^2}$ in the range of $0< r<1.8$ fm.
Fitting parameters at $t/a=14$ are shown in Table~\ref{tab-fit},
and corresponding fitting results are shown by red bands in Fig.~\ref{Fig-V}.

\begin{figure}[htbp]
    \centering
    \includegraphics[width=8.0cm]{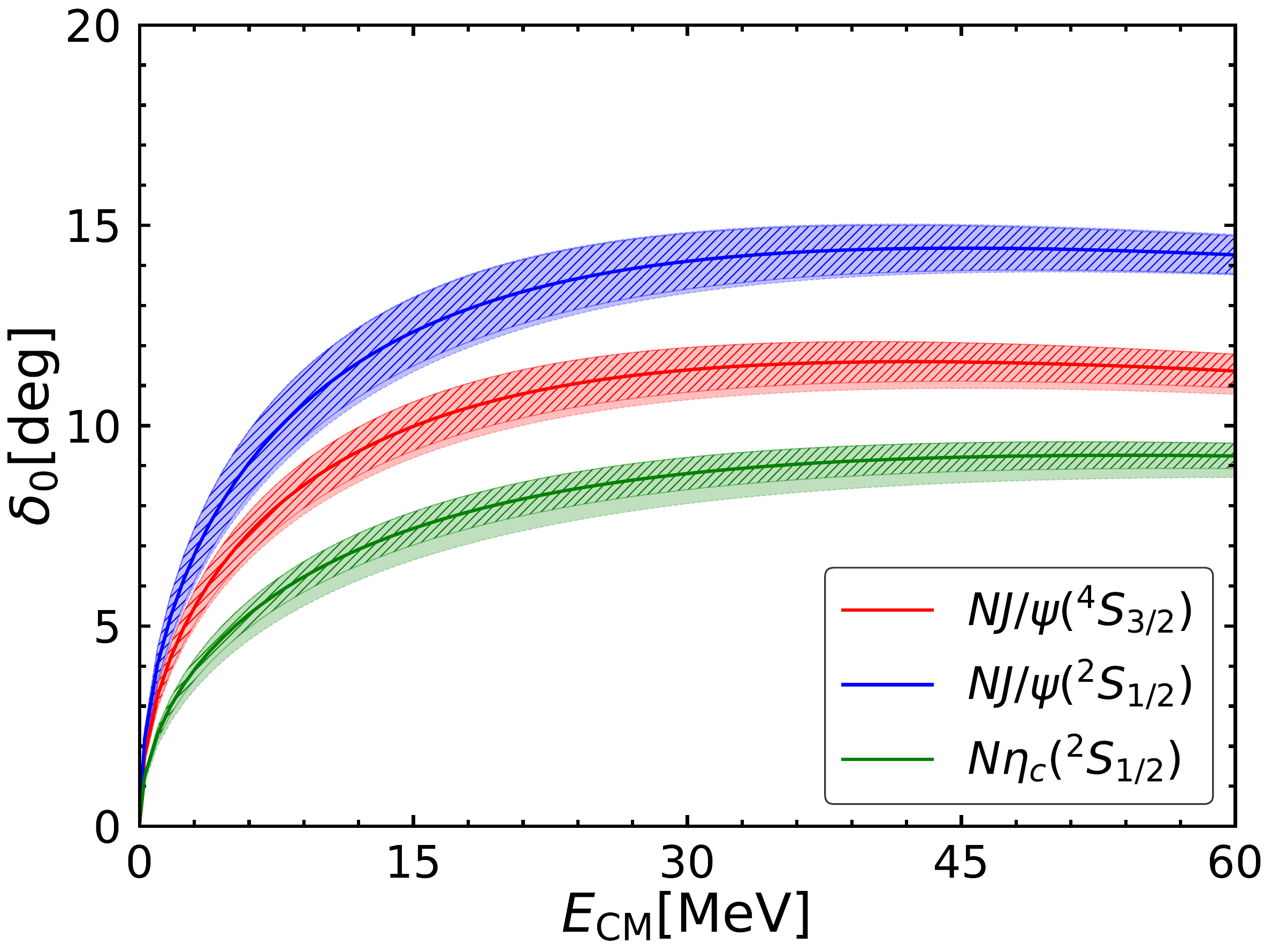}
    \caption{The $N$-$c\bar c$ scattering phase shifts. The central value and statistical error obtained with $V_\text{fit}(r)$ at $t/a=14$ are shown by solid lines and inner bands. The outer bands show total uncertainty obtained by adding statistical error and systematic error in quadrature,
    }
    \label{Fig-delta}
\end{figure}

Shown in Fig.~\ref{Fig-delta} is the $N$-$c\bar c$ scattering phase shifts calculated with $V_\text{fit}(r)$
as a function of center of mass kinetic energy $E_\text{CM}=\sqrt{m_N^2+k^2}+\sqrt{m_{c\bar c}^2+k^2}-m_N-m_{c\bar c}$.
The central value and statistical error shown by solid lines and inner bands are
obtained by solving the Schr\"odinger equation with $V_\text{fit}(r)$ at $t/a=14$,
while the total uncertainties shown by outer bands are obtained by adding the statistical error in quadrature with the systematic error estimated by comparing results at different $t/a$. No bound state is found for all three $N$-$c\bar c$ channels.

Scattering phase shifts near the threshold can be expanded as,
\begin{align}
    k\cot\delta_0=\frac{1}{a_0}+\frac12 r_\text{eff}k^2+O(k^4),
\end{align}
where $a_{0}$ and $r_\text{eff}$ are the scattering length and the effective range.
In Table~\ref{tab-scattering}, we show the $a_{0}$ and $r_\text{eff}$ for low-energy $N$-$c\bar c$ scattering obtained from the scattering phase shifts in Fig.~\ref{Fig-delta}.

 \begin{table}[htbp]
\begin{center}
\caption{Fit parameters with statistical errors in the parentheses at $t/a=14$. }
\begin{tabular}{lccc}
  \hline\hline
    ~~~~~~&~~$N$-$J/\psi(^4S_{3/2})$~~~&~~$N$-$J/\psi(^2S_{1/2})$~~~&~~$N\eta_c(^2S_{1/2})$\\
  \hline
$a_1$ [MeV] &$51(1)$ &  $101(1)$ & $264(14)$ \\
$a_2$ [MeV] &$13(6)$ &  $33(6)$ &  $28(13)$ \\
$a_3$ [MeV] &$22(5)$ & $23(8)$ & $22(2)$ \\
$b_1$ [fm] &$0.09(1)$ & $0.13(1)$ & $0.11(1)$ \\
$b_2$ [fm] &$0.49(7)$ & $0.44(5)$ & $0.24(6)$ \\
$b_3$ [fm] &$0.82(6)$ & $0.83(9)$ & $~0.77(3)$ \\
\hline
$\chi^2/\text{dof}$ & $0.4$ & $0.4$ & $0.6$\\
\hline\hline
\end{tabular} \label{tab-fit}
\end{center}
\end{table}   

 \begin{table}[htbp]
\begin{center}
\caption{The $N$-$c\bar c$ scattering length $a_0$ and effective range $r_\text{eff}$ with statistical error ($1$st parentheses) and systematic error ($2$nd parentheses).}
\begin{tabular}{lll}
  \hline\hline
    ~~~channel~~~~~~~~&~~~$a_0$ [fm]~~~~~~~~~&~~~$r_\text{eff}$ [fm]\\
  \hline
$N$-$J/\psi(^4S_{3/2})$ ~~~~~~~~~&$0.30(2)\left(^{+0}_{-2}\right)$ ~~~~~~~~~~&$3.25(12)\left(^{+6}_{-9}\right)$ \\
$N$-$J/\psi(^2S_{1/2})$ ~~~~~~~~~&$0.38(4)\left(^{+0}_{-3}\right)$ ~~~~~~~~~~&$2.66(21)\left(^{+0}_{-10}\right)$\\
$N\eta_c(^2S_{1/2})$ ~~~~~~~~~&$0.21(2)\left(^{+0}_{-1}\right)$ ~~~~~~~~~~&$3.65(19)\left(^{+0}_{-6}\right)$ \\
\hline\hline
\end{tabular} \label{tab-scattering}
\end{center}
\end{table}   

The scattering length given in Table~\ref{tab-scattering} has a direct phenomenological application, the mass shift of $J/\psi$ meson in nuclear medium can be well constrained by our present results.
According to QCD sum rule studies in Refs.~\cite{Koike1997,Hatsuda1995,Hayashigaki:1998ey},
the $J/\psi$ meson mass reduction in nuclear matter under the linear density approximation is given by,
\begin{align}
    \delta m_{J/\psi}\simeq \frac{2\pi(m_N+ m_{J/\psi})}{m_Nm_{J/\psi}}a^\text{spin-av}_{J/\psi} \rho_\text{nm}=19(3)~\text{MeV},
\end{align}
where $\rho_\text{nm}=0.17~\text{fm}^{-3}$ denotes the normal nuclear matter density, and the $N$-$J/\psi$ spin-averaged scattering length is defined as $a^\text{spin-av}_{NJ/\psi}=\frac{2a^\text{(3/2)}_0+a^\text{(1/2)}_0}{3}$.

\subsection{Comparison of scattering lengths}
The previous studies on the $N$-$c\bar c$ scattering lengths by various approaches
were scattered in a large range:
(i) Data of the $J/\psi$ photoproduction off the proton are used to constrain the corresponding amplitude constructed by assuming the vector meson dominance~\cite{Gryniuk:2016mpk, Pentchev:2020kao} and by enforcing the low-energy unitary~\cite{JPAC:2023qgg}, which give rise to $a^\text{spin-av}_{NJ/\psi}$ around $O(1\sim10)\times10^{-3}$~fm and $O(1)$~fm, respectively.
(ii) Previous lattice studies report $a^\text{spin-av}_{NJ/\psi}$ to be
$0.7(5)$~fm~\cite{Yokokawa:2006td}, $0.33(5)$~fm~\cite{Kawanai:2010ru} under quenched approximation, $0.47(1)$~fm~\cite{Sugiura:2019pye} with $m_\pi=700$~MeV, and consistent with zero within large uncertainties by Refs~\cite{Liu:2008rza,Skerbis:2018lew}.
(iii) The QCD multiple expansion method complemented with low-energy QCD theorem
is adopted in Refs~\cite{Kaidalov:1992hd, Brodsky:1997gh, Sibirtsev:2005ex}, and resulting scattering lengths range from $0.05~$ fm to $0.37$ fm or more due in part to large uncertainties in the chromo-polarizability~$\alpha_{J/\psi}$~\cite{Polyakov:2018aey}. 
(iv) Using QCD sum rule, Ref~\cite{Hayashigaki:1998ey} predicts $a^\text{spin-av}_{NJ/\psi}=0.10(2)$ fm.
(v) Within a coupled channel framework including $\Lambda_c\bar{D}^{(*)}$, Ref~\cite{Du:2020bqj} finds $a^\text{spin-av}_{NJ/\psi}$ is around $O(1)\times10^{-3}$ fm.
These scattering lengths in one way or another surfer from either badly fixed parameters in corresponding models
or lacking of some important dynamics of QCD in corresponding calculations, which also underline the importance of performing first-principles QCD calculations with (almost) physical quark masses as what we did in the present study.

\section{Summary and conclusions}
In summary, we present a realistic study on the low-energy interaction
between a nucleon ($N$) and a charmonium ($J/\psi $ and $\eta _{c}$) based
on ($2+1$)-flavor lattice QCD calculations with nearly physical light quark
masses ($m_{\pi}=146$ MeV) and physical charm quark mass. The $N$-$c
\bar c$ potential derived from the corresponding hadronic correlation function
is found to be attractive at all distances, and its long-range ($0.9< r<1.8$~fm)
part is identified to be consistent with the two-pion exchange interaction.
The resulting scattering lengths are
$0.30(2)\left (^{+0}_{-2}\right )$ fm,
$0.38(4)\left (^{+0}_{-3}\right )$ fm, and
$0.21(2)\left (^{+0}_{-1}\right )$ fm for $N$-$J/\psi $ with spin
$3/2$, with spin $1/2$, and $N$-$\eta _{c}$ with spin $1/2$, respectively. These results
are orders-of-magnitude larger than those inferred from $J/\psi $ photoproduction
off the proton assuming the vector meson dominance, and are largely improved
compared with previous lattice calculations under the quenched approximation
and at heavy pion mass. The results presented here could be checked by
future measurements of the femtoscopic correlations of $p$-$J/\psi $, which
have been shown to be sensitive to the underlying interaction~\cite{Krein:2020yor,Krein:2023azg}.

To further improve our results, we are currently in progress to perform
physical-point calculations with the ($2+1$)-flavor lattice QCD configurations
generated by the HAL QCD Collaboration~\cite{Aoyama:2024cko}. Also, it
is interesting to examine the long-range potential at even larger distances
by substantially increasing statistics, as well as its universality across
other hadron pairs. Further results will be reported elsewhere.

\section*{Note added}
After the completion of the present paper, we noted a recent study~\cite{Wu:2024xwy} which suggests the low-energy $N$-$J/\psi$ scattering is dominated by soft gluon exchange mechanism.

\section*{Acknowledgments}
We thank members of the HAL QCD Collaboration for stimulating discussions.
We thank members of the PACS Collaboration for the gauge configuration generation conducted on the K computer at RIKEN.
The lattice QCD measurements have been performed on Fugaku and HOKUSAI supercomputers at RIKEN.
We thank ILDG/JLDG \cite{ldg}, which serves as essential infrastructure in this study.
This work was partially supported by HPCI System Research Project (hp120281, hp130023, hp140209, hp150223, hp150262, hp160211, hp170230, hp170170, hp180117, hp190103, hp200130, hp210165, hp210212, hp220240, hp220066, hp220174, hp230075, hp230207, hp240157 and hp240213), the JSPS (Grant Nos. JP18H05236, JP22H00129, JP19K03879, JP21K03555, and JP23H05439),   
``Program for Promoting Researches on the Supercomputer Fugaku'' (Simulation for basic science: from fundamental laws of particles to creation of nuclei) 
and (Simulation for basic science: approaching the new quantum era)
(Grants No. JPMXP1020200105, JPMXP1020230411), 
and Joint Institute for Computational Fundamental Science (JICFuS).
YL, TD and TH weresupported by Japan Science and Technology Agency (JST) as part of Adopting Sustainable Partnerships for Innovative Research Ecosystem (ASPIRE), Grant Number JPMJAP2318.









\end{document}